\documentclass[twoside]{article}
\usepackage{tikz}
\usepackage{amsmath}
\usepackage{color, hyperref}
\usepackage[style=apa, backend=biber, maxcitenames=2, sorting=nyt, hyperref=true]{biblatex}
\usetikzlibrary{positioning, arrows.meta, shapes.geometric, calc, fit}
\usepackage{mathrsfs,amsthm,amssymb,amsfonts, bm,mathtools, microtype, bbm, times, epsfig, comment, graphicx, booktabs,siunitx, subfig, algorithm, algpseudocode, multirow, makecell, float}
\usepackage{lmodern}
\usepackage{newtxtext}

\newcommand{\size}[2]{\scalebox{#1}{$\displaystyle #2$}}
\def\E{\mathbb{E}}
\def\R{\mathbb{R}}
\def\P{\mathbb{P}}
\def\L{\size{1.2}{\mathscr{L}}}
\newcommand{\sfrac}[2]{\raisebox{0.8ex}{$#1$}\!/\!\raisebox{-0.8ex}{$#2$}}
\newcommand{\boldscr}[1]{\text{\ooalign{$\mathscr{#1}$\cr\kern.7pt$\mathscr{#1}$}}}

\definecolor{darkblue}{RGB}{0,51,102}
\definecolor{graynode}{RGB}{230,230,230}
\definecolor{lightgray}{RGB}{200,200,200}
\definecolor{fancypink}{RGB}{88,10,211}
\definecolor{steelblue}{rgb}{0.27, 0.51, 0.71}
\definecolor{lblue}{rgb}{0.17, 0.41, 0.91}
\definecolor{greenblue}{rgb}{0.0, 0.4, 0.6} 
\definecolor{redblue}{rgb}{0.2, 0.33, 0.71}   
\definecolor{seablue}{rgb}{0.13, 0.55, 0.73}
\usepackage[accepted]{aistats2025}
\theoremstyle{plain}
\newtheorem{theorem}{Theorem}[section]
\newtheorem{proposition}[theorem]{Proposition}

\theoremstyle{definition}
\newtheorem{assumptioncounter}{Assumption}[section]
\newtheorem{assumption}[assumptioncounter]{Assumption}

\theoremstyle{remark}

\begin{document}

\twocolumn[

\aistatstitle{Feed-Forward Panel Estimation for Discrete-time Survival Analysis of Recurrent Events with Frailty}

\aistatsauthor{Borna Bateni \And Peyman Bateni \And Bishwadeep Bhattacharyya \And Devin Reeh}

\aistatsaddress{ Department of Statistics\\  UCLA\\  Los Angeles, CA \And  \\    \And  Department of Statistics\\  UCLA\\  Los Angeles, CA\And  Department of Statistics\\  UCLA\\  Los Angeles, CA}]

\begin{abstract}
 In recurrent survival analysis where the event of interest can occur multiple times for each subject, frailty models play a crucial role by capturing unobserved heterogeneity at the subject level within a population. Frailty models traditionally face challenges due to the lack of a closed-form solution for the maximum likelihood estimation that is unconditional on frailty. In this paper, we propose a novel method: \textbf{F}eed-\textbf{F}orward \textbf{P}anel estimation for discrete-time \textbf{Surv}ival Analysis (\textbf{FFPSurv}). Our model uses variational Bayesian inference to sequentially update the posterior distribution of frailty as recurrent events are observed, and derives a closed form for the panel likelihood, effectively addressing the limitation of existing frailty models. We demonstrate the efficacy of our method through extensive experiments on numerical examples and real-world recurrent survival data. Furthermore, we mathematically prove that our model is identifiable under minor assumptions.
\end{abstract}

\section{\MakeUppercase{Introduction}}
\label{submission}

Survival analysis provides powerful statistical tools for determining how different variables impact the time until an event of interest occurs. Survival models are extensively used in statistics, econometrics, social sciences, and medicine, such as to estimate the duration of unemployment \parencite{lancaster1979}, predicting the criminal recidivism of individuals released from incarceration \parencite{schmidt1989predicting}, or evaluating the impact of prognostic variables in patient outcomes such as death or cancer recurrence \parencite{katzman2018deepsurv, gillick2001guest}.

\begin{figure}[ht]
\centering
\resizebox{\columnwidth}{!}{
\begin{tikzpicture}[node distance=1.5cm and 1.0cm, auto, thick, 
    box/.style={
        rectangle, 
        draw=darkblue, 
        fill=gray!30, 
        rounded corners=3pt, 
        minimum height=0.6cm, 
        minimum width=1cm, 
        text centered,
        shading=axis, 
        shading angle=135, 
        top color=gray!20, 
        bottom color=gray!50
    },
    boxy/.style={
        rectangle, 
        draw=steelblue, 
        fill=gray!10, 
        rounded corners=3pt, 
        minimum height=0.6cm, 
        minimum width=1cm, 
        text centered,
        shading=axis, 
        shading angle=135, 
        top color=gray!10, 
        bottom color=gray!30
    },
    boxl/.style={
        rectangle, 
        draw=lblue, 
        fill=gray!5, 
        rounded corners=3pt, 
        minimum height=0.6cm, 
        minimum width=1cm, 
        text centered,
        shading=axis, 
        shading angle=135, 
        top color=gray!5, 
        bottom color=gray!10
    },
    boxop/.style={
    rectangle, 
    draw=red!70, 
    fill=yellow!20,
    thick, 
    rounded corners=5pt, 
    text centered,
    font=\large\boldmath, 
    shading=axis, 
    shading angle=135, 
    top color=yellow!10, 
    bottom color=yellow!30 
    },    
    boxn/.style={
        rectangle, 
        draw=seablue, 
        fill=gray!05, 
        rounded corners=3pt, 
        minimum height=0.6cm, 
        minimum width=1cm, 
        text centered,
        shading=axis, 
        shading angle=135, 
        top color=gray!05, 
        bottom color=gray!10
    },
    boxd/.style={
        rectangle, 
        draw=greenblue, 
        fill=gray!10, 
        rounded corners=3pt, 
        minimum height=0.6cm, 
        minimum width=1cm, 
        text centered,
        shading=axis, 
        shading angle=135, 
        top color=gray!10, 
        bottom color=gray!30
    },
    boxx/.style={
        rectangle, 
        draw=redblue, 
        fill=gray!10, 
        rounded corners=3pt, 
        minimum height=0.6cm, 
        minimum width=1cm, 
        text centered,
        shading=axis, 
        shading angle=135, 
        top color=gray!10, 
        bottom color=gray!30
    },
    boxt/.style={rectangle, draw=black, rounded corners=5pt, double, thick, minimum height=0.9cm, minimum width=1.4cm, text centered},
    process/.style={rectangle, draw=black, dashed, thick, minimum height=0.9cm, text centered},
    likboxt/.style={rectangle, draw=darkblue,rounded corners=5pt,  minimum height=0.9cm, minimum width=2.0cm, text centered, font=\small},
    bayes/.style={rectangle, draw=black, dotted, thick, text centered},
    every node/.append style={font=\footnotesize}
]

    \node[boxy] (y1) {\( y_{i,1} \)};
    \node[boxd, below= 0.1cm of y1] (d1) {\( d_{i,1} \)};
    \node[boxx, below= 0.1cm of d1] (x1) {\( x_{i,1} \)};
    \node[boxt, fit=(y1) (d1) (x1), inner sep=3pt, label=below:{}] (group1) {};
    
    \node[boxy, right=1.2cm of y1] (y2) {\( y_{i,2} \)};
    \node[boxd, below= 0.1cm of y2] (d2) {\( d_{i,2} \)};
    \node[boxx, below= 0.1cm of d2] (x2) {\( x_{i,2} \)};
    \node[boxt, fit=(y2) (d2) (x2),  inner sep=3pt, label=below:{}] (group2) {};

    \node[boxy, right=1.2cm of y2] (y3) {\( y_{i,3} \)};
    \node[boxd, below= 0.1cm of y3] (d3) {\( d_{i,3} \)};
    \node[boxx, below= 0.1cm of d3] (x3) {\( x_{i,3} \)};
    \node[boxt, fit=(y3) (d3) (x3),  inner sep=3pt, label=below:{}] (group3) {};

    \node[boxy, right=1.2cm of y3] (yJ) {\( y_{i,J} \)};
    \node[boxd, below= 0.1cm of yJ] (dJ) {\( d_{i,J} \)};
    \node[boxx, below= 0.1cm of dJ] (xJ) {\( x_{i,J} \)};
    \node[boxt, fit=(yJ) (dJ) (xJ),  inner sep=3pt, label=below:{}] (groupJ) {};
    
    \node[boxn, below right=0.8cm and 0.2cm of group1] (nu1) {\( \size{.9}{\nu\big|[y\ d\ x]_{i,1}} \)};
    \node[boxn, below right=0.8cm and 0.2cm of group2] (nu2) {\( \size{.9}{\nu\big|[y\ d\ x]_{i,1:2}} \)};
    \node[boxn, below right=0.8cm and 0.2cm of group3] (nu3) {\( \size{.9}{\nu\big|[y\ d\ x]_{i,1:3}} \)};
    \node[boxn, below right=0.8cm and 0.2cm of groupJ] (nuJ) {\( \size{.9}{\nu\big|[y\ d\ x]_{i,1:J}} \)};
    \node[boxn, left=0.3cm of nu1] (nu) {\( \size{.9}{\nu\sim\Gamma_{\alpha,\kappa}} \)};

    \node[boxl, below left=1.95cm and 1.8cm of nu2] (L1) {\( \mathscr{L}([y\ d\ x]_{i,1}) \)};
    \node[boxl, below left=2.75cm and 0.12cm of nu2] (L2) {\( \mathscr{L}([y\ d\ x]_{i,2}|[y\ d\ x]_{i,1}) \)};
    \node[boxl, below left=3.55cm and -.48cm of nu2] (L3) {\( \mathscr{L}([y\ d\ x]_{i,3}|[y\ d\ x]_{i,1:2}) \)};
    \node[boxl, below left=4.35cm and -1.29cm of nu2] (LJ) {\( \mathscr{L}([y\ d\ x]_{i,J}|[y\ d\ x]_{i,1:J\!-\!1}) \)};

    \node[bayes, fit=(group1)(group2)(group3)(groupJ), minimum width=3cm, label=left:{\textbf{Data}}, draw=fancypink, very thick] (datat) {};

    \node[bayes, fit=(nu)(nu1)(nu2)(nu3)(nuJ), minimum width=3cm, label=left:{\textbf{Frailty}}, draw=fancypink, very thick] (frailtyt) {};

\node[bayes, fit=(L1)(L2)(L3)(LJ), minimum width=3cm, 
    label=left:{\parbox{0.8cm}{\textbf{Bayes}\\\textbf{Chain}}}, 
    draw=fancypink, very thick] (bayesboxt) {};
    
    \node[likboxt, right=1.2cm of bayesboxt, align=center, fill=graynode!40] (max) {\( \size{1.2}{\underset{\phi,\delta,\alpha,\kappa}{\text{Max}}\mathscr{L}([y\ d\ x]_{i,1:J})}\)};
    
    \draw[->, thick] (group1.east) -- (group2.west);
    \draw[->, thick] (group2.east) -- (group3.west);
    \draw[->, dashed] (group3.east) -- (groupJ.west);
    
    \draw[->, thick, bend right=25] (group1.south) to (nu1);
    \draw[->, thick, bend right=25] (group2.south) to (nu2);
    \draw[->, thick, bend right=25] (group3.south) to (nu3);
    \draw[->, thick, bend right=25] (groupJ.south) to (nuJ);

    \draw[->, thick] (nu) -- (nu1);
    \draw[->, thick] (nu1) -- (nu2);
    \draw[->, thick] (nu2) -- (nu3);
    \draw[->, thick, dashed] (nu3) -- (nuJ);

\pgfdeclarelayer{background}
\pgfsetlayers{background,main}

\begin{pgfonlayer}{background}
    \coordinate (commonY) at ($(nu.south)!0.15!(L1.north)$);
    \coordinate (commonY2) at ($(commonY) + (1.75,0)$);
    \coordinate (commonY3) at ($(commonY) + (3.5,0)$);
    \coordinate (commonYJ) at ($(commonY) + (4.5,0)$);
    \draw[-, thick] ($(group1.south) + (-0.185, 0)$) |- (commonY) -| ($(L1.north) + (0.8,0)$);
    \draw[-, thick] ($(group2.south) + (-0.605, 0)$) |- (commonY2) -| ($(L2.north) + (1.47,0)$);
    \draw[-, thick] ($(group3.south) + (-0.52, 0)$) |- (commonY3) -| ($(L3.north) + (1.52,0)$);
    \draw[-, thick] ($(groupJ.south) + (-0.52, 0)$) |- (commonYJ) -| ($(LJ.north) + (1.75,0)$);
    
    \draw[->, thick] (nu.south) |- (commonY) -| ($(L1.north) + (.8,0)$);
    \draw[->, thick] (nu1.south) |- (commonY2) -| ($(L2.north) + (1.47,0)$);
    \draw[->, thick] (nu2.south) |- (commonY3) -| ($(L3.north) + (1.52,0)$);
    \draw[->, thick] (nu3.south) |- (commonYJ) -| ($(LJ.north) + (1.75,0)$);
\end{pgfonlayer}
\draw[->, very thick] (bayesboxt) -- (max);
\node[boxop, below left=.59cm and -1.63cm of nu2, align=center, fill=graynode!40] (intbox) {\( \ \  \size{.71}{\size{1}{\int} \mathscr{L}([y\ d\ x]\ |\ \nu)\ d\text{F}(\nu)} \ \  \)};
\node[boxop, right=.25cm of bayesboxt, align=center, fill=graynode!40] (prodbox) {\( \size{.71}{\prod} \)};
\end{tikzpicture}}
    \caption{An overview of \textbf{FFPSurv}. Observed variables time ($y$), censor ($d$) and features ($\mathbf{x}$) at each occurrence are used to update the frailty ($\nu$). Conditional likelihood terms are obtained by integrating across frailty. Then, the likelihood terms are merged into a single joint likelihood using the Bayes chain which is maximized to estimate parameters.}
    \label{fig:enter-label}
\end{figure}
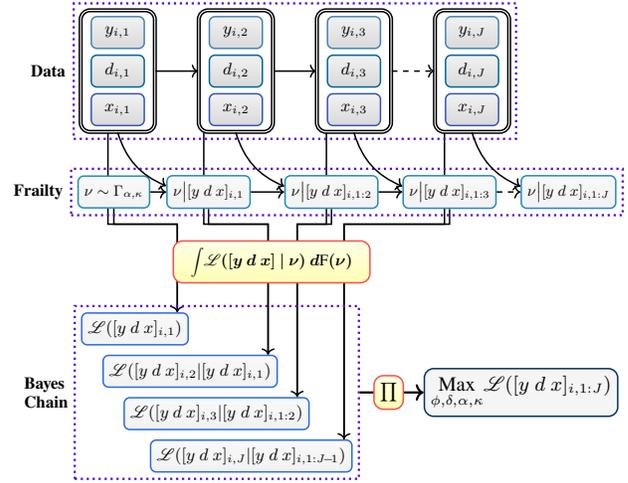
Existing survival models can be broadly grouped into three categories: 1) parametric models that assume specific distributions for survival times, offering parameter efficiency but requiring accurate distributional assumptions; 2) nonparametric models that make minimal distributional assumptions about survival time, providing flexibility but typically requiring larger sample sizes; and 3) semi-parametric models that incorporate a parametric and a nonparametric component. Notably, the \textcite{cox1972} Proportional Hazards model (\textit{CoxPH}) was the first semi-parametric model to be implemented in survival analysis. \textit{CoxPH} incorporated regression-like parameters associated with explanatory variables into an arbitrary and unknown function of time, later named the \textit{baseline hazard function}. Accordingly, the model derived a \textit{proportional hazard} for every subject, demonstrating the probability of the event occurring given that it has not yet occurred.

A key focus within survival analysis is on models that address the occurrence of recurrent events, where the event of interest can happen repeatedly over time, rather than being a terminal occurrence. Examples include unemployment spells or recurrence of non-terminal diseases. This area explores the concept of \textbf{frailty} \parencite{beard1959note,vaupel1979impact,Duch08}, the subject-level unobserved heterogeneity that impacts the risk of the occurrence of the event. The incorporation of frailty transforms the data into a panel structure, as the time-to-event outcomes observed from the same subject will become correlated when the frailty effect is unobserved. 

Although many existing survival models are designed to handle continuous time-to-event data, in most practical scenarios, especially in economics and social sciences, duration data is typically captured as discrete and arranged into intervals of a predetermined length. This presents complications especially when the discrete data showcases a multitude of `ties'; instances where survival times are identical. The continuous model framework disproportionately weights these ties, interpreting them as instances with the same survival times; while in reality, they are merely survival times grouped within the same interval. This has given rise to a branch of \textit{discrete survival models} concerned with these `grouped' survival times \parencite{moffitt1985, sueyoshi1995, gensheimer2019scalable}.


Among discrete survival models is that of \textcite{hanhausman1990} which handles grouped duration data by partitioning the area under the curve of the baseline hazard into increments equal to the grouped intervals and estimating each increment as a separate parameter. While Han and Hausman's model offers a straightforward closed-form solution, it assumes the events are independent across the data, and thus falls short of formulating the inter-dependencies between the multiple occurrences when modeling recurrent events. In this paper, we expand on the work of Han and Hausman, to propose an improved model for discrete-time survival analysis designed for recurrent events.

\textbf{Our contributions} are as follows:
\begin{itemize}
    \item We propose \textbf{FFPSURV}, a longitudinal discrete-time model designed for subjects with unobserved frailty who could potentially undergo multiple duration spells of awaiting an event.
    \item We present mathematical proof that \textbf{FFPSURV} is identifiable under minor assumptions.
    \item We present experimental evidence that demonstrates the empirical efficacy of \textbf{FFPSURV} as compared to existing baselines in both simulated and real-world data applications.
\end{itemize}

\section{\MakeUppercase{Related Work}}

The concept of unobserved heterogeneity influencing the time-to-event of every subject was long present in actuarial sciences. \textcite{makeham1867law}  linked the hazard rate with a constant (equivalent to the baseline hazard) and a progression in time coefficient that can vary across the population. Using Makeham’s law, \textcite{beard1959note} provided an interpretation for the diverse mortality rates, and coined the term ‘longevity factor’ to denote the varying coefficient of the hazard rate, serving to classify the population. \textcite{vaupel1979impact} were first to use the term `frailty' to explain how frail people have an unobserved factor that contributes to a heightened morbidity rate.

Contemporary research in survival analysis with frailty includes models focused on covariate effect quantification by computing the panel likelihood across multiple duration outcomes for each subject. Methods include the Expectation Maximization (EM) algorithm \parencite{nielsen1992, ripatti2002maximum}, Penalized likelihood maximization \parencite{therneau2003penalized, ripatti2000estimation}, Hierarchical likelihood approaches \parencite{ha2001}, and distribution-less pseudo-likelihood maximization \parencite{gorfine2006}. Other approaches include estimating time-to-event data quantiles which has found widespread application, notably within econometrics \parencite{harding2012quantile, chen2019quantile}. 

Another focus is on predictive survival analysis for recurrent events, which has seen a notable shift towards models that excel in non-linear pattern recognition. This includes methods like random survival forests \parencite{ishwaran2008random,chen2012}, and models that use a deep neural network structure \parencite{faraggi1995neural, katzman2018deepsurv}. To the best of our knowledge, the only neural network-based survival models specifically designed for recurrent data analysis include \textit{CRESA} \parencite{gupta2019cresa} and \textit{DeepPAMM} \parencite{kopper2022deeppamm}. \footnote{It is important to distinguish between models addressing recurrent outcomes and those using a recurrent neural network (RNN) architecture. While some studies use RNNs for survival data \parencite{ren2019deep, lee2019dynamic, nagpal2021deep}, they focus on terminal events only. In these cases, "recurrent" refers to the model architecture, not the event type.} \textit{CRESA} uses an LSTM-based architecture to learn the distributions of time-to-event and competing events directly, offering predictions on the timing and cause of events through its LSTM network and cause-specific sub-networks. \textit{DeepPAMM} leverages Piecewise Exponential Additive Models (\textit{PAM}s) and integrates deep neural networks to enhance its predictive capacity. It approximates the hazard function through an additive model that incorporates both linear and non-linear effects of covariates. 

Our model situates itself alongside the innovations for maximum likelihood estimation, proposing an alternative approach based on the variational Bayesian chain rule for sequentially computing the discrete-time panel likelihood in the presence of unobserved frailty. We provide numerical evidence that the estimation approach is comparable to the state-of-the-art models in terms of covariate estimation. We provide two variants of our model, one with linear and the other with a neural network-based feature transformation, and show that they both are comparable to the existing predictive models in terms of predictive accuracy.

\section{\MakeUppercase{Methodology}}\label{sec:meth}

\subsection{Problem Definition}

We consider a sample of $n$ subjects, denoted as $i=1,\dots,n$ where each subject $i$ can undergo multiple spells of time-to-event outcomes for a recurrent event, denoted as $\textbf{t}_i = \left[t_{i1},\dots, t_{i\size{.6}{{J}_{i}}}\right]^\top$. For each subject, an observed $p$-dimensional vector of independent features $\mathbf{x}_{ij} = \left[x_{ij1},\dots, x_{ijp}\right]^\top$ at each outcome $j$ is provided, and an unobserved Gamma-distributed frailty parameter $\nu_i\sim \Gamma_{\alpha,\kappa}$ where $\alpha, \kappa$ are the shape and rate parameters of the Gamma distribution, respectively.

We adopt a mixed proportional hazards setup, where the hazard function for the $j^{\text{th}}$ outcome of the $i^{\text{th}}$ individual is the product of the unobserved heterogeneity $\nu_i$, a non-parametric baseline hazard function $\lambda_0:\R^{+}\to \R^{+}$, and a finite-parametric non-negative transformation function $\Phi:\R^p\to\R^+$ of the features $\mathbf{x}_{ij}$:
\begin{align*}
    \lambda_{ij}(t|\size{0.8}{\nu_i})\coloneq&\size{.75}{\lim\limits_{\Delta t\to 0^+}\frac{1}{\Delta t}}\cdot{\P\left(t_{ij}\in[t,t+\Delta t)\mid t_{ij}\geq t, \size{0.8}{\mathbf{x}_{ij}, \nu_i}\right)}\\ = & \ \nu_i\cdot\lambda_0(t)\cdot\Phi(\mathbf{x}_{ij})
\end{align*}
Consequently, a conditional cumulative distribution function (CDF) of the outcome, given the frailty is obtained:
$$F_{ij}(t|\size{.8}{\nu_i})=1-\exp\left(-\nu_i\cdot\Phi(\mathbf{x}_{ij})\cdot\int_0^{t}\lambda_0(\zeta)d\zeta\right)$$
And the following unconditional CDF is derived by integrating out the latent frailty:
\begin{equation}\label{eq:0}\begin{aligned}F_{ij}(t)&=\int_0^\infty F_{ij}(t|\size{.8}{\nu_i})d\text{F}(\nu_i)\\ &= 1-\left(1+\frac{1}{\kappa}\cdot\Phi(\mathbf{x}_{ij})\cdot\int_0^{t}\lambda_0(\zeta)d\zeta\right)^{-\alpha}\end{aligned}\end{equation}

Next, we adapt the framework to analyze discrete time-to-event outcomes. Consider a ``grouping" process that partitions each outcome into intervals of length $\psi$. We define $\tau_{ij}=\psi\cdot\lfloor \frac{t_{ij}}{\psi}\rfloor$ as the discretized outcome for the $i^{\text{th}}$ subject at the $j^{\text{th}}$ spell. Moreover, consider a right-censoring process and let $\varsigma_{ij}$ denote a randomly distributed censoring time with a CDF denoted by $G_{ij}$. The censoring time is also discretized by the same grouping process and the discretized censoring time is denoted by $c_{ij}=\psi\cdot\lfloor \frac{\varsigma_{ij}}{\psi}\rfloor$. Lastly, let $d_{ij}=\mathbbm1_{\tau_{ij}<c_{ij}}$ be the censoring indicator, and $y_{ij} = \min(c_{ij},\tau_{ij})$ be the final observed outcome outcome. For each subject $i$, we observe the set of triplets  $\left[\mathbf{y}_i, \mathbf{d}_i, \mathbf{X}_i\right] = \left\{\left[ y_{ij}, d_{ij}, \mathbf{x}_{ij}\right]\right\}_{j=1}^{J_i}$, where $\mathbf{y}_i=\left[y_{i1},\dots, y_{i\size{.6}{{J}_{i}}}\right]^\top$ and $\mathbf{d}_i=\left[d_{i1},\dots, d_{i\size{.6}{{J}_{i}}}\right]^\top$ are the vectors for the observed outcomes and censoring indicators, and $\mathbf{X}_i = \left[\mathbf{x}_{i1},\dots, \mathbf{x}_{i\size{.6}{{J}_{i}}}\right]^\top$ is the matrix of features.

It is possible to construct a closed-form expression for the likelihood of the triplet $\left[y_{ij}, d_{ij},\mathbf{x}_{ij}\right]$, using the CDF for the continuous outcome $F_{ij}$ as derived in \ref{eq:0}, and the censoring CDF $G_{ij}$. The following proposition outlines the likelihood.
\begin{proposition}
\label{prop:prop0}
The likelihood function of each triplet of observed outcomes $\left[y_{ij}, d_{ij},\mathbf{x}_{ij}\right]$ can be constructed as follows:
\begin{align*}
    \L_{ij}(y,1) &= \P(y_{ij} = y, d_{ij}=1\mid \mathbf{x}_{ij}) \\&= \left(1-G_{ij}(y+\psi)\right).\left(F_{ij}(y+\psi)-F_{ij}(y)\right) \\ & \\
    \L_{ij}(y,0) &= \mathbb{P}(y_{ij} = y, d_{ij}=0\mid \mathbf{x}_{ij}) \\&= \left(1-F_{ij}(y)\right).\left(G_{ij}(y+\psi)-G_{ij}(y)\right)
\end{align*}
\end{proposition}

Further, we assume that the censoring process is ``non-informative'' \parencite{Liang95}, thus allowing us to drop the contribution of the censoring distribution to the likelihood and construct a unified form for the likelihood of $\left[y_{ij}, d_{ij},\mathbf{x}_{ij}\right]$ as:

\begin{equation} \label{eq:1}
\begin{aligned}
&\size{1}{\L_{ij}(y,d)}=\size{.8}{\bigg(}F_{ij}(y+\psi)-F_{ij}(y)\size{.8}{\bigg)^{d}}\cdot\size{.78}{\bigg(}1-F_{ij}(y)\size{.8}{\bigg)^{1-d}} \\
&= \size{.8}{\bigg(1+\frac{\Phi(\mathbf{x}_{ij})}{\kappa}\cdot\!\int_0^y\!\! \lambda_0(\zeta)d\zeta\bigg)^{\!\!-\alpha}}\!\!\!\! - d\cdot\size{.8}{\bigg(1+\frac{\Phi(\mathbf{x}_{ij})}{\kappa}\cdot\!\int_0^{y+\psi}\!\!\! \lambda_0(\zeta)d\zeta\bigg)^{\!\!-\alpha}}    
\end{aligned}
\end{equation}

 Next, we examine a discretized version of the baseline hazard function. Let $y^*\coloneq \size{.8}{\underset{i,j}{\max}}\ \ y_{ij}$ denote the highest observed outcome among all subjects and spells. We partition the cumulative baseline hazard in the interval $[0,y^*)$ into equal increments $\delta_k$ with lengths equal to $\psi$ and define the discretized baseline hazard as the sum of the increments:
 \begin{align*}
     \delta_k&\coloneq\int_{(k-1).\psi}^{k.\psi}\lambda_0(\zeta)d\zeta\ \ ;\ \ k\in\left\{1,2,\dots,\sfrac{y^*}{\psi}\right\}
     \\ \lambda_0^{\text{dis.}}(t)&\coloneq \sum_{k=0}^{\size{.6}{\sfrac{y^*}{\psi}}} \delta_k\!\cdot\! \mathbbm{1}_{t\size{.6}{\in}\left(\size{.6}{(\!k\!-\!1\!)}\cdot\psi,\size{0.6}{k}\cdot\psi\right]}\ +\ \delta^*\!\!\cdot\!\mathbbm{1}_{t>y^*}
 \end{align*}
Where $\delta^*=\delta_{\size{.5}{\sfrac{y^*}{\psi}}+1}\coloneq \int_{y^*}^{\infty}\lambda_0(\zeta)d\zeta$ represents the remaining portion of the integral that falls outside of the discretized increments, and $\delta_0=0$. Note that, unlike the continuous baseline hazard function which is inherently non-parametric, it is possible to have a finite-dimensional sufficient statistic $\left\{\hat\delta_1,\dots,\hat\delta_r\right\}$ for the discretized baseline hazard, where $r = \left|\left\{k\in\{1,\dots,\sfrac{y^*}{\psi}\}\mid \delta_k>0\right\}\right|$.

We plug in the discretized baseline hazard into the likelihood derived in \ref{eq:1} to obtain a discretized version of the likelihood:
\begin{equation}\label{eq:2}\begin{aligned}
    \size{.8}{\L_{ij}^{\text{dis.}}(y,d)}&= \size{.85}{\bigg(1+\frac{\Phi(\mathbf{x}_{ij})}{\kappa}\cdot\sum_{\zeta=0}^{\size{.6}{\sfrac{y}{\psi}}} \delta_\zeta\bigg)^{\!\!-\alpha}\!\!\!\!\!-d\!\cdot\!\bigg(1+\frac{\Phi(\mathbf{x}_{ij})}{\kappa}\cdot\sum_{\zeta=0}^{\size{.6}{\sfrac{y}{\psi}}+1} \delta_\zeta\bigg)^{\!\!-\alpha}}
\end{aligned}\end{equation}
It is notable that in the present framework, $\mathscr{L}_{ij}^{\text{dis.}}$ is not merely an approximation of $\mathscr L_{ij}$, but is equivalent to it. Let $\mathcal{S}_{\size{.6}{yd}}=\{k\cdot\psi|k\in\mathbb N_0\}\times\{0,1\}$ be the space of all possible values for the pair $[y_{ij},d_{ij}]$. Define $\mathcal{F}=2^{{\mathcal{S}_{\size{.4}{yd}}}}$ as the power set of $\mathcal{S}_{\size{.6}{yd}}$, and consider the measures $\mu(y,d) = \mathscr L_{ij}(y,d)\cdot \mathbbm{1}_{y\leq y^*}$ and $\mu^{\size{.6}{\text{dis.}}}(y,d)= \mathscr L^{\size{.6}{\text{dis.}}}_{ij}(y,d)\cdot \mathbbm{1}_{y\leq y^*}$ over $\mathcal F$. One can readily observe that for any set $A\in\mathcal{F}$, the Radon-Nikodym derivative of the two measures $\frac{d\mathcal{\mu}}{d\mathcal{\mu}^{\text{dis.}}}({A})$ equals $1$. Hence, the continuous and discrete likelihoods are equivalent over the panel of observed data, and moving forward we will adopt equation \ref{eq:2} as the expression for the likelihood in our analysis.
\subsection{Estimation of Panel Likelihood}
A significant challenge considering the estimation of the panel likelihood for the triplet $\left[\mathbf{y}_i, \mathbf{d}_i, \mathbf{X}_i\right]$ is that integrating out the latent frailty does not yield a closed-form expression for the unconditional panel likelihood.  Here we introduce a novel framework for the panel likelihood of discrete survival outcomes using a variational Bayesian network.

Let $\boldscr{L}_i(\mathbf{y},\mathbf{d})$ denote the panel likelihood for $\left[\mathbf{y}_i, \mathbf{d}_i, \mathbf{X}_i\right]$. Using the Bayes chain rule, the panel likelihood can be written as the product of conditional likelihoods for each observation, given all previous observations:
\begin{equation}\label{eq:2.5}\begin{aligned}
    \size{.95}{\size{1.2}{\boldscr{L}}_i(\mathbf{y}_i,\mathbf{d}_i)=\L\!\!{_{i1}}(y_{i1},d_{i1})\cdot\prod_{j=2}^{J_i}\L\!\!{_{ij}}\left(y_{ij},d_{ij}\bigg| \big[y_{i\ell},d_{i\ell}\big]_{\ell=1}^{j-1}\right)}
\end{aligned}\end{equation}
Note that while different observed spells from the same subject are independent under the fully specified model, when the frailty term is latent the spells become marginally correlated, and the conditional likelihood can be obtained by integrating out the posterior distribution of the latent frailty:
\begin{equation}\label{eq:2.75}\begin{aligned}
    \size{.85}{\L\!\!{_{ij}}\!\!\left(\!y_{ij},d_{ij}\bigg| \big[y_{i\ell},d_{i\ell}\big]_{\ell=1}^{j-1}\!\right)\!=\!\!\int_{0}^\infty\!\!\!\!\! \L\!\!{_{ij}}\!\!\left(y_{ij},d_{ij}\big| \nu_i\right)\!\!\cdot \!d\text{F}\!\left(\!\nu_i\size{.9}{\bigg|}\big[y_{i\ell},d_{i\ell}\big]_{\ell=1}^{j-1}\!\right)}
\end{aligned}\end{equation}

We now turn to the problem of identifying a conjugate distribution for the frailty variable. When having continuous outcomes, the Gamma distribution serves as an exact conjugate, facilitating the derivation of a closed-form panel likelihood \parencite{murphy95}. However, such ease is absent in the discrete case. Thus, we employ variational inference methods to derive a pseudo-conjugate frailty distribution.

Let the Gamma distribution with hyperparameters $\alpha$ and $\kappa$ for shape and rate be the prior distribution for the frailty parameter, $\nu_i\sim\Gamma_{\alpha,\kappa}$. Using the Bayes rule, we obtain:
\begin{equation}\label{eq:4}\begin{aligned}
    &\text{F}\left(\nu_i\big|y_{ij},d_{ij}\right) = \frac{\L_{ij}\left(y_{ij},d_{ij}\big|\nu_i\right)}{\L_{ij}\left(y_{ij},d_{ij}\right)}\cdot \text{F}(\nu_i)\\
    &=\!\dfrac{\size{.75}{\exp\!\bigg(\!-\!\nu_i\!\cdot\!{\Phi(\mathbf{x}_{ij})}\cdot\!\!\sum_{\zeta=0}^{\frac{y_{ij}}{\psi}} \delta_\zeta\bigg)\!-d_{ij}\exp\!\bigg(\!-\!\nu_i\!\cdot\!{\Phi(\mathbf{x}_{ij})}\cdot\!\!\!\!\sum_{\zeta=0}^{\frac{y_{ij}}{\psi}+1} \!\!\delta_\zeta\bigg)}}{\size{.75}{\bigg(1+\frac{\Phi(\mathbf{x}_{ij})}{\kappa}\cdot\sum_{\zeta=0}^{\frac{y_{ij}}{\psi}} \delta_\zeta\bigg)^{\!\!-\alpha}\!\!\!\!\!-d_{ij}\!\cdot\!\bigg(1+\frac{\Phi(\mathbf{x}_{ij})}{\kappa}\cdot\sum_{\zeta=0}^{\frac{y_{ij}}{\psi}+1}\!\! \delta_\zeta\bigg)^{\!\!-\alpha}}}\!\cdot\! \frac{\size{.85}{\nu_{i}^{\alpha-\!1}\cdot e^{\!-\kappa \nu_i}}}{\size{.75}{\Gamma(\alpha)\cdot \kappa^{\!-\alpha}}}\\
    &= \frac{\size{1}{\nu_i^{\alpha-\!1}\cdot\left(e^{\!-\nu_i\cdot\xi_{ij}}-d_{ij}\!\cdot\!e^{\!-\nu_i\cdot\xi_{ij}'}\right)}}{\size{1}{\Gamma(\alpha)\cdot\left(\xi_{ij}^{-\alpha}-\ d_{ij}\!\cdot\!\xi_{ij}'^{-\alpha}\right)}}
\end{aligned}\end{equation}

where $$\xi_{ij} \coloneq \kappa+\Phi(\mathbf{x}_{ij})\sum_{\zeta=1}^{\frac{y_{ij}}{\psi}}\delta_\zeta\quad ;\quad \xi_{ij}' \coloneq \kappa+\Phi(\mathbf{x}_{ij})\sum_{\zeta=1}^{\frac{y_{ij}}{\psi}+1}\!\!\delta_\zeta$$

Denote by $\epsilon_{ij}\coloneq \frac{\xi_{ij}'-\xi_{ij}}{\xi_{ij}'}$ the relative difference between the terms $\xi_{ij}$ and $\xi_{ij}'$. Inherently, $\epsilon_{ij}$ resides in the open unit interval $(0,1)$ and asymptotically converges to $0$ as $\frac{\xi_{ij}}{\xi_{ij}'}\to 1$, and to $1$ as $\frac{\xi_{ij}}{\xi_{ij}'}\to 0$. Based on the value of $\epsilon_{ij}$ and the censoring status, we propose the subsequent three variational inference techniques aimed at approximating the posterior frailty distribution with a Gamma distribution. Starting with the case of a censored outcome ($d_{ij}=0$), we can directly derive an exact posterior Gamma distribution. Subsequently, as $\epsilon_{ij}$ approaches its extremities, the posterior distribution demonstrates weak convergence to a Gamma distribution. Finally, for all other instances, we apply the method of moment matching to approximate the posterior distribution as a Gamma distribution.
\begin{proposition}
\label{prop:prop1}
When $d_{ij}=0$ the posterior frailty will be exactly Gamma distributed:
$$\text{F}\left(\nu_i\big|y_{ij},d_{ij}=0\right)=\Gamma\size{.8}{\left(\alpha,\xi_{ij}\right)}$$
\end{proposition}

\begin{proposition}
\label{prop:prop2}
When $d_{ij}=1$ and $\epsilon_{ij}$ is near its asymptotic values, the posterior frailty will converge in distribution to a Gamma distribution:
\begin{align*}\text{F}\left(\nu_i\big|y_{ij},d_{ij}=0\right)&\stackrel{d}{\longrightarrow}\Gamma\size{.8}{(\alpha\!+\!1,\size{.8}{\frac{\xi_{ij}+\xi_{ij}'}{2}})}\quad\  \text{as}\ \ \ \epsilon_{ij}\to0\\
\text{F}\left(\nu_i\big|y_{ij},d_{ij}=0\right)&\stackrel{d}{\longrightarrow}\Gamma\size{.8}{\left(\alpha,\xi_{ij}\right)}\quad \quad\quad\quad\text{as}\ \ \ \epsilon_{ij}\to1\end{align*}
\end{proposition}

\begin{proposition}
\label{prop:prop3}
For any arbitrary $\epsilon_{ij}$ and when $d_{ij}=1$, the moment matching approximant Gamma distribution of the posterior frailty will be:
\begin{align*}
    &\text{F}\left(\nu_i\big|y_{ij},d_{ij}=0\right)\simeq \Gamma\size{.8}{(\tilde\alpha, \tilde\kappa)}\quad\quad\text{; where:}\\
    &\tilde\alpha=\alpha\ \cdot\!\size{.8}{\left({\frac{(1-(1-\epsilon_{ij})^{\alpha+1})^2}{(1-(1-\epsilon_{ij})^{\alpha})(1-(1-\epsilon_{ij})^{\alpha+2})-\alpha\cdot(1-\epsilon_{ij})^\alpha(\epsilon_{ij})^2}}\right)}\\
    &\tilde\kappa={\xi_{ij}\!\cdot\!\size{.8}{\left(\frac{(1-(1-\epsilon_{ij})^{\alpha})(1-(1-\epsilon_{ij})^{\alpha+1})}{(1-(1-\epsilon_{ij})^{\alpha})(1-(1-\epsilon_{ij})^{\alpha+2})-\alpha\cdot(1-\epsilon_{ij})^\alpha(\epsilon_{ij})^2}\right)}}
\end{align*}
\end{proposition}

Within the framework outlined above, the Gamma distribution emerges as an approximate conjugate distribution for the frailty variable, allowing the derivation of a closed-form expression for the posterior frailty distribution after observing multiple outcomes from a subject. To operationalize this, define functions $f^{{}^{\alpha}}_{ij}(.)$ and $f^{{}^{\kappa}}_{ij}(.)$ based on the above propositions, that update the shape and rate parameters of the Gamma distribution, respectively. Accordingly, the frailty distribution parameters after observing multiple outcomes will sequentially update into multi-layered decompositions of the functions:
\begin{align*}
    \text{F} \!\left(\!\nu_i\size{.9}{\bigg|}\big[y_{i\ell},d_{i\ell}\big]_{\ell=1}^{j-1}\!\right) \simeq \Gamma\!\size{.8}{\left(\tilde\alpha^{(j-1)},\tilde\kappa^{(j-1)}\right)}
\end{align*}
where: \begin{align*}
    \size{.8}{\tilde \alpha^{(j-1)}=f^{{}^{\alpha}}_{i,j-1}(f^{{}^{\alpha}}_{i,j-2}(\dots f^{{}^{\alpha}}_{i1}(\alpha)))\ };\size{.8}{\ \tilde\kappa^{(j-1)}=f^{{}^{\kappa}}_{i,j-1}(f^{{}^{\kappa}}_{i,j-2}(\dots f^{{}^{\kappa}}_{i1}(\kappa)))}
\end{align*}
Incorporating the posterior frailty into equation \ref{eq:2.75}, we achieve a closed form for the conditional likelihood of the triplet $\left[y_{ij}, {d}_{ij}, \mathbf{x}_{ij}\right]$ given previous observations:
\begin{align*}
    &\size{.95}{\L\!\!{_{ij}}\!\!\left(\!y_{ij},d_{ij}\bigg| \big[y_{i\ell},d_{i\ell}\big]_{\ell=1}^{j-1}\!\right)\!=}\\
    &\size{.85}{\bigg(1+\frac{\Phi(\mathbf{x}_{ij})}{\tilde\kappa^{(j-1)}}\cdot\sum_{\zeta=0}^{{\frac{y_{ij}}{\psi}}} \delta_\zeta\bigg)^{\!\!-\tilde\alpha^{(j-1)}}\!\!\!\!\!\!\!\!-d_{ij}\!\cdot\!\bigg(1+\frac{\Phi(\mathbf{x}_{ij})}{\tilde\kappa^{(j-1)}}\cdot\sum_{\zeta=0}^{{\frac{y_{ij}}{\psi}}+1} \delta_\zeta\bigg)^{\!\!-\tilde\alpha^{(j-1)}}}
\end{align*}
Finally, by applying this formulation to equation \ref{eq:2.5}, we successfully derive a closed form for the panel likelihood. The panel likelihood can then be maximized over its parameters, $\big\{\Phi(.),\{\delta_1,\dots,\delta_r\},\alpha,\kappa\big\}$ enabling the derivation of the Maximum Likelihood Estimates.
\subsection{The Choice of the Features Transformation Function}
The conventional approach within the \textit{CoxPH} framework models the features transformation function an an exponential transformation of a linear combination of features, \(\Phi(\mathbf x _{ij}) = \exp({\mathbf x_{ij}\!\cdot\!\beta})\) where \(\beta\) represents the vector of coefficients. While such  a framework offers interpretability and simplicity, it may not well reflect the complexities and potential nonlinear interactions within the features. In response to that, recent studies have proposed various neural network structures to replace the linear combination \parencite{faraggi1995neural, katzman2018deepsurv, Kvamme2019, Hu2023}. In such cases, a neural network structure will be plugged into the exponential transformation, $\Phi(\mathbf{X}_{ij}) = \exp(g(\mathbf{X}_{ij};\theta))$ with $\theta$ being the weights of the neural network. 

In the current study, we propose two distinct approaches to model $\Phi(.)$; the first approach explores a traditional linear model emphasizing interpretability and explanatory analysis, while the second approach employs a neural network, potentially offering enhanced predictive power in scenarios with intricate data patterns. Both approaches will be rigorously evaluated and compared to state-of-the-art competing models in terms of predicting power and estimating accuracy.

\subsection{Sensitivity to the Frailty Distribution}\label{sec:sens}

Although the Gamma distributional family has been the most popular choice for the frailty distribution in applications \parencite{van2001duration}, its selection and the sensitivity of survival models to this distributional assumption have been subjects of debate in the survival analysis literature. Some arguments challenge the suitability of the Gamma distribution. Notably, \textcite{heckman84b} argue that the maximum likelihood estimator of the frailty should yield a discrete rather than continuous distribution under certain regularity conditions, and propose an EM algorithm for estimating the discrete support and associated probabilities. 

However, other research offers support in favor of the Gamma distribution. \textcite{abbring2007unobserved} prove that the posterior distribution of frailty converges to a Gamma distribution whenever the prior distribution is regularly varying at zero—a condition met by a broad class of priors. Also, \textcite{aalen1992modelling} proves that for the compound Poisson family of frailty models, the frailty distribution among subjects who do not observe the event is a mixture of a Gamma distribution with a probability mass at zero. These results provides a theoretical rationale for the Gamma distribution's widespread adoption in frailty models. Additionally, empirical studies indicate that survival data often lacks sufficient information to precisely identify the shape of the frailty distribution \parencite{ kortram1995constructive, horowitz1999semiparametric}. Consequently, the choice of the frailty distribution, such as Gamma, is unlikely to be contradicted or strongly influenced by data, making it a practical and robust choice in most applications.

Our own simulations, presented in Subsection \ref{sec:sim1}, offer additional empirical support for the robustness of the Gamma distributional assumption. Specifically, in the third simulation setup, we demonstrate that \textbf{FFPSurv} remains robust even when the underlying frailty distribution deviates significantly from the Gamma assumption—in this case, adopting a discrete distribution with two possible support values.

\subsection{Computational Complexity}\label{sec:compx}

Similar to  the \textit{CoxPH} model, \textbf{FFPSurv} exhibits linear scaling with the number of features \(p\) during both training and hazard prediction, resulting in an overall computational complexity of \(\mathcal{O}(p)\) in standard survival settings. However, when considering recurrent event settings, \textbf{FFPSurv} introduces an additional layer of complexity. Specifically, because the gradient calculations in \textbf{FFPSurv} condition on each prior event, the complexity becomes \(\mathcal{O}(J \cdot p)\), where \(J\) is the (maximal) number of prior outcomes across all subjects. In contrast, \textit{CoxPH} maintains its complexity as \(\mathcal{O}(p)\), when dealing with recurrent events.

\section{\MakeUppercase{Model Identifiability}}\label{sec:iden}
\subsection{Preventing Overparametrization:}
We start deriving identifiability conditions for the maximum likelihood estimates by positing the necessary assumptions for the well-parametrization of the optimization problem. Although $\Phi(.),\alpha,\kappa$ provide a fixed number of parameters for estimation regardless of the number of observations, the number of the parameters $\{\delta_1,\dots,\delta_r\}$ is not fixed and is relevant to the maximum observed outcome $y^*$ and the grouping interval length $\psi$. In extreme cases where $\psi$ is too small and the observed outcomes $y_{ij}$ tend to overshoot, the number of $\delta_i$'s could grow faster than the number of observations, making the model overparametrized and impossible to identify. In order to avoid overparametrization, we need to make the subsequent assumptions.
\begin{assumption}\label{as:0}
The inverse of the squared length of the grouping interval is dominated by the number of observed individuals:\begin{align*}
    \psi^{-2}={o}(n)
\end{align*}  
\end{assumption}
\begin{assumption}\label{as:1}
The unconditional distribution of the duration outcomes has finite moments. Hence, For all $i=1,\dots,n$ and $j=1,\dots,J_i$:
\begin{align*}
    &\E\left(y_{ij}^k\right)<\infty \quad\quad \text{ for $k\in\R^+$}
\end{align*}
\end{assumption}
\begin{theorem}\label{theo:0}
    Under assumptions \ref{as:0} and \ref{as:1}, the number of the estimating parameters will be asymptotically dominated by the number of observations; that is, the number of $\delta_i$'s is bounded from above by $\sfrac{y^*}{\psi}$, and: $$\frac{\sfrac{y^*}{\psi}}{n}\xrightarrow[]{\P}0$$ 
\end{theorem}
\subsection{Uniqueness of Parameter Estimations}
We now present an adaptation of the general identification framework for duration models with multiple outcomes \parencite{honore93, heckman84, elbers1982true} to the discrete-time setup of our model and demonstrate the uniqueness of the maximum likelihood estimators. To achieve this, we need the following assumptions:
\begin{assumption}
\label{as:3}
The baseline hazard function $\lambda_0(\cdot)$ exhibits the properties of a proper hazard function; namely, it is a non-negative, locally integrable function satisfying
$$\int_\tau^t \lambda_0(\zeta)d\zeta\xrightarrow[]{t\to\infty}\infty\quad\quad \text{for all }\tau\in[0,\infty)$$
The implication of this assumption for the discrete case is that each of the increments $\left\{ \delta_1,\dots,\delta_{\size{.55}{\sfrac{y^*}{\psi}}} \right\}$ reside in $[0,\infty)$, and that $\delta^* = \int_{\size{.55}{\sfrac{y^*}{\psi}}}^\infty \lambda_0(\zeta)d\zeta=\infty$.
 \end{assumption}
\begin{assumption}
 \label{as:4}
For all duration intervals in which no outcomes have been observed, the corresponding hazard increment $\delta_\zeta$ is assigned to zero. These parameters, along with $\delta^*=\infty$, are treated as known and therefore not requiring estimation. Accordingly, we have: $$\forall \delta_\zeta \ \text{such that }\ \size{.9}{\left\{y_{ij}\big| y_{ij}\!\in\! \big[(\zeta\!-\!1)\!\cdot\! \psi\ ,\ \zeta\!\cdot\!\psi\big)\right\}=\emptyset}:\quad \delta_\zeta=0$$
\end{assumption}
\begin{assumption}\label{as:5}
The space of possible features $\mathcal S(\mathbf x)$ includes at least two points, $\mathbf x_1$ and  $\mathbf x_2$, such that $\Phi(\mathbf x_1)=1$ and $\Phi(\mathbf x_2)\neq 1$.
\end{assumption}
 \begin{theorem}
\label{theo:2}
     Under Assumptions \ref{as:3}--\ref{as:5}, there exists a one-to-one mapping from the domain of the parameters $\left\{\Phi(\cdot), \{\delta_1,\dots,\delta_r\},\alpha,\kappa\right\}$ to the data, which makes the model estimates uniquely identifiable.
 \end{theorem}

\section{\MakeUppercase{Simulations}}\label{sec:sim}
In this section, we detail simulation studies conducted to evaluate the performance of our models and benchmark them against a variety of state-of-the-art models. We introduce two variants of our estimating method: \textbf{{FFPSURV-L}} which employs the (exponentiated) linear combination of features for the feature transformation function $\Phi(\cdot)$, and \textbf{{FFPSURV-N}} which uses a neural network structure.
\subsection{Feature Coefficient Estimation Performance}\label{sec:sim1}
We evaluate the performance of the linear variant of our model, \textbf{{FFPSURV-L}}, in estimating the feature coefficients across four distinct setups. The \textbf{first setup} involves simulated data for $n=250$ subjects, each having $J=4$ duration spells, using a Gamma-distributed latent frailty, $\nu\sim \Gamma_{1,1}$, and a logarithmic baseline hazard function, $\lambda_0(t) = \log(1\! +\! c{t})$. The next setups all maintain the core structure of the first setup, each with variations only in a specific aspect. The \textbf{second setup} modifies the baseline hazard function into a more complex pattern, $\lambda_0(t)=c_1\sqrt{ t}\cdot \sin^2\!\left(c_2t\right)$. The \textbf{third setup} tests the model's robustness to frailty distribution mis-specification by employing a discrete frailty distribution, $\P(\nu) = 0.5\cdot\!\mathbbm{1}_{\nu=0.5}+0.5\cdot\!\mathbbm{1}_{\nu=5}$. The \textbf{fourth setup} explores the performance in a wider dataset, changing the data dimensions to $n=50$ subjects each with $J=20$ outcomes.

Duration outcomes are generated according to the proportional hazard setup, $\lambda(t) = \nu\!\cdot\!\lambda_0(t)\!\cdot\!\Phi(x)$ with a linear feature transformation, $\Phi(\mathbf x) = \exp(\mathbf x\beta)$, where $\beta$ is a vector of size $3$ with an oracle value equal to $[0.4,-1,1]$. Outcomes are then grouped in intervals of unit length, $y=\lfloor t\rfloor$, and right-censored at $y_{\max}=80$. 

The competing models for this section are Cox Proportional hazard 
 with unobserved shared frailty (\textit{{CoxPH-F}}) \parencite{gorfine2006, monaco2018general}, the H-likelihood frailty estimation (\textit{{HLfrail}}) \parencite{ha2001, ha2017} and the expectation maximization frailty estimation (\textit{{EMfrail}}) \parencite{balan2019}.

 Table \ref{tab:0} presents the aggregated mean and standard deviation of coefficient estimates for the four models, derived from $300$ bootstrap iterations. While all models perform well, the aggregated mean performance of our model consistently ranks as the highest, indicating competitiveness with the state-of-the-art survival methodology in terms of coefficient estimates.

\begin{table}[h]
\caption{Numerical Comparison of Coefficient Estimates.}
\label{tab:0}
\vskip 0.15in
\begin{center}
\footnotesize
\setlength{\tabcolsep}{3pt} 
\resizebox{\columnwidth}{!}{ 
\begin{sc}
\begin{tabular}{lccc}
\hline
{} & ${\beta_1}$ & $\beta_2$ & $\beta_3$ \\
\hline
\hline
\multicolumn{4}{c}{$\size{.8}{\text{Setup 1}}$} \\
\hline
\hline
CoxPH-F & 0.297 \tiny(0.234) & -0.897 \tiny(0.230) & 1.012 \tiny(0.096) \\
HLfrail & 0.387 \tiny(0.190) & -0.956 \tiny(0.194) & 0.952 \tiny(0.088) \\
EMfrail & 0.391 \tiny(0.192) & -0.971 \tiny(0.196) & 0.966 \tiny(0.090) \\
\textbf{FFPSURV-L} & \textbf{0.406} \tiny(0.199) & \textbf{-1.010} \tiny(0.205) & \textbf{1.005} \tiny(0.095) \\

\hline
\hline
\multicolumn{4}{c}{$\size{.8}{\text{Setup 2}}$} \\
\hline
\hline
CoxPH-F & {0.292} \tiny(0.247) & -0.925 \tiny(0.240) & {1.008} \tiny(0.088) \\
HLfrail & 0.375 \tiny(0.196) & -0.919 \tiny(0.187) & 0.920 \tiny(0.077) \\
EMfrail & 0.382 \tiny(0.197) & -0.933 \tiny(0.191) & 0.937 \tiny(0.078) \\
\textbf{FFPSURV-L} & \textbf{0.408} \tiny(0.212) & \textbf{-1.001} \tiny(0.204) & \textbf{1.003} \tiny(0.084) \\
\hline
\hline
\multicolumn{4}{c}{$\size{.8}{\text{Setup 3}}$} \\
\hline
\hline
CoxPH-F & 0.361 \tiny(0.168) & {-0.755} \tiny(0.218) & {1.016} \tiny(0.099) \\
HLfrail & 0.380 \tiny(0.167) & -0.952 \tiny(0.201) & 0.942 \tiny(0.077) \\
EMfrail & 0.380 \tiny(0.169) & -0.950 \tiny(0.202) & 0.939 \tiny(0.078) \\
\textbf{FFPSURV-L} & \textbf{0.404} \tiny(0.179) & \textbf{-1.012} \tiny(0.216) & \textbf{1.001} \tiny(0.084) \\
\hline
\hline
\multicolumn{4}{c}{$\size{.8}{\text{Setup 4}}$} \\
\hline
\hline
CoxPH-F & 0.382 \tiny(0.182) & -0.972 \tiny(0.203) & 0.986 \tiny(0.104) \\
HLfrail & 0.380 \tiny(0.170) & -0.964 \tiny(0.180) & 0.979 \tiny(0.073) \\
EMfrail & {0.380} \tiny(0.170) & {-0.963} \tiny(0.180) & {0.978} \tiny(0.073) \\
\textbf{FFPSURV-L} & \textbf{0.391} \tiny(0.175) & \textbf{-0.992} \tiny(0.186) & \textbf{1.005} \tiny(0.075)\\
\hline
Oracle & {0.4}\quad & {-1}\quad & {1}\quad\\
\hline
\end{tabular}
\end{sc}}

\end{center}
\vskip -0.1in
\end{table}

\subsection{Predictive Performance with Complex Data Patterns}

We benchmark the predictive ability of the \textbf{FFPSURV} models across a non-linear data generating process. To that end, we adapt the initial data generation configuration from setup 1 in the previous subsection, and modify the feature transformation function to non-linear pattern $\Phi(\mathbf{x}) = \exp\big(\beta_1\cdot x_1 + \beta_2\cdot (x_2+c_2)^2 + \log(\beta_3\cdot x_3+c_3) + \beta_4\cdot x_4\big)$. We generate a dataset comprising $n=300$ subjects each with $4$ observations, with the train-test split being $2/3$ to $1/3$. For this section the competing models are the widely used semi-parametric \textit{CoxPH}, the parametric Accelerated Failure Time (\textit{AFT}) model with Weibull, Log-Normal, and Log-Logistic distributions for the baseline hazard, and the deep neural nets survival models \textit{DeepSurv} \parencite{katzman2018deepsurv} and \textit{DeepPAMM} \parencite{kopper2022deeppamm}. Except for \textit{DeepPAMM} which is originally designed to handle recurrent events data, for the competing models, we adopted a one-hot encoding strategy for subject ID's to accommodate the recurrent event structure. 

Results across $100$ bootstrap iterations are presented in Table \ref{tab:1}, underscoring the competitive predictive accuracy of both the linear \textbf{FFPSURV-L} and neural network \textbf{FFPSURV-N} variants of our model relative to the established benchmarks. The neural network variant, despite its complexity, does not markedly outperform the linear variant of the model.

\begin{table}[h]
\caption{Predictive Ability of Models on Simulated Data.}
\label{tab:1}
\vskip 0.15in
\begin{center}
\footnotesize 
\setlength{\tabcolsep}{3pt} 
{ 
\begin{sc}
\begin{tabular}{lcccc}
\hline
{} & C-Index $\uparrow$  \\
\hline
DeepPAMM                & 0.6166 \tiny \textpm 0.0295   \\
Weibull AFT             & 0.6190 \tiny \textpm 0.0214   \\
Log-Norm AFT            & 0.6213 \tiny \textpm 0.0193   \\
Log-Logis AFT           & 0.6188 \tiny \textpm 0.0215   \\
DeepSurv                & 0.5990 \tiny \textpm 0.0316   \\
CoxPH                   & 0.6201 \tiny \textpm 0.0199   \\
\textbf{FFPSURV-L}      & \textbf{0.6194} \tiny \textpm \textbf{0.0198}  \\
\textbf{FFPSURV-N}      & \textbf{0.6226} \tiny \textpm \textbf{0.0195}  \\
\hline
\end{tabular}
\end{sc}
}
\end{center}
\vskip -0.1in
\end{table}

\section{\MakeUppercase{Evaluation on Real Data}} \label{sec:real}

We further evaluate \textbf{FFPSURV-L} on 3 datasets involving real recurrent survival events that may take place multiple times.

The Unemployment insurance data \parencite{harding2012quantile}, collected and published by the Upjohn Institute, consists of 5015 observations of the duration of unemployment, with the time at which the unemployment spell ends as the response variable of interest, for 3329 individuals between the ages of 20 and 65 from 1997 to 2001. In addition to the length of the unemployment spell, each observation has additional features including whether the worker received unemployment benefits, the age of the worker at the beginning of the unemployment spell, whether the worker had permanent employment before unemployment, and the state where the worker collects her benefits.

The Staph dataset \parencite{pammtools1, pammtools2, pammtools3, pammtools4} is a subset of the Drakenstein Child Health Study that focuses on the time until staphylococcus aureus infection in children, with possible recurrence. The dataset includes 374 observations from 137 children with a maximum of 6 recurrences. The features of each observation consist of an enumeration of health events and an indicator variable describing whether the mother of the child tested positive for HIV.

The Bladder dataset \parencite{survival-package, survival-book} consists of observations of the recurrences of bladder cancer in 85 individuals with the maximum number of recorded recurrences capped at 4. The features of each observation consist of the type of treatment (Placebo, Pyridoxine (vitamin B6), or Thiotepa), the initial number of tumors, the size of the largest initial tumor, and a measure of how many recurrences have previously happened.

In each setting, we randomly sample 80\% of observations for training with the remaining 20\% used for testing. We used stratified sampling to ensure that the same ratio of observed and censor samples is achieved in both splits. We repeat our evaluation 10 times with different random train and test splits and report the averaged results.
\begin{table}[h]
\caption{Predictive Ability of Models on Real Data.}
\label{tab:2}
\vskip 0.15in
\begin{center}
\footnotesize 
\setlength{\tabcolsep}{3pt} 
\resizebox{\columnwidth}{!}{ 
\begin{sc}
\begin{tabular}{lcc}
\toprule
\textbf{Unemployment Insurance} & C-Index $\uparrow$ & IBS $\downarrow$ \\
\midrule
DeepPAMM                & 0.564 \tiny \textpm 0.032 & -- \\
Weibull AFT             & 0.585 \tiny \textpm 0.024 & 0.297 \tiny \textpm 0.035 \\
Log-Normal AFT          & 0.592 \tiny \textpm 0.023 & 0.299 \tiny \textpm 0.032 \\
Log-Logistic AFT        & 0.595 \tiny \textpm 0.023 & \textbf{0.293 \tiny \textpm 0.032} \\
DeepSurv                & 0.584 \tiny \textpm 0.026 & -- \\
CoxPH                   & 0.598 \tiny \textpm 0.018 & 0.341 \tiny \textpm 0.027 \\
\textbf{FFPSURV-L}                  & \textbf{0.601 \tiny \textpm 0.016} & 0.370 \tiny \textpm 0.029 \\
\midrule
\textbf{Staph} & C-Index $\uparrow$ & IBS $\downarrow$ \\
\midrule
DeepPAMM                & 0.591 \tiny \textpm 0.059 & -- \\
Weibull AFT             & 0.544 \tiny \textpm 0.099 & 0.291 \tiny \textpm 0.036 \\
Log-Normal AFT          & 0.544 \tiny \textpm 0.101 & 0.302 \tiny \textpm 0.034 \\
Log-Logistic AFT        & 0.546 \tiny \textpm 0.099 & 0.298 \tiny \textpm 0.034 \\
DeepSurv                & 0.657 \tiny \textpm 0.067 & -- \\
CoxPH                   & 0.661 \tiny \textpm 0.075 & 0.268 \tiny \textpm 0.025 \\
\textbf{FFPSURV-L}                  & \textbf{0.663 \tiny \textpm 0.078} & \textbf{0.262 \tiny \textpm 0.025} \\
\midrule
\textbf{Bladder} & C-Index $\uparrow$ & IBS $\downarrow$ \\
\midrule
DeepPAMM                & 0.574 \tiny \textpm 0.103 & -- \\
Weibull AFT             & 0.582 \tiny \textpm 0.091 & 0.338 \tiny \textpm 0.129 \\
Log-Normal AFT          & 0.587 \tiny \textpm 0.085 & \textbf{0.336 \tiny \textpm 0.131} \\
Log-Logistic AFT        & 0.591 \tiny \textpm 0.089 & 0.337 \tiny \textpm 0.128 \\
DeepSurv                & 0.585 \tiny \textpm 0.083 & -- \\
CoxPH                   & 0.599 \tiny \textpm 0.096 & 0.382 \tiny \textpm 0.092 \\
\textbf{FFPSURV-L}                  & \textbf{0.604 \tiny \textpm 0.096} & 0.370 \tiny \textpm 0.095 \\
\bottomrule
\end{tabular}
\end{sc}}
\end{center}
\vskip -0.1in
\end{table}

The results are shown in Table \ref{tab:2}. For all models, we report the C-index for baseline hazard prediction on the test set. Furthermore, where available, we also calculate the integrated brier score for predicted survival from day 1 up to the maximum even occurrence reported in the dataset. As demonstrated, \textbf{FFPSURV} achieves the top mean c-index across all 3 benchmarks, while achieving the top integrated brier score on the Staph dataset.

\section{\MakeUppercase{Discussion}}

In this paper, we have introduced an improved method for recurrent discrete-time panel survival analysis. Our model is able to produce a closed form panel likelihood estimate by sequentially updating frailty after each occurrence of the event for a subject. We have demonstrated the improved performance of our method over existing baselines in a range of simulated and real world recurrent survival event environments. In the future, further exploration for improved feature transformation architectures may provide fruitful avenues for improved performance in survival modelling. In addition, exploring the impact of modelling multiple recurrent inter-dependent events that are shared by a single subject may further improve efficacy in areas where events can be highly interconnected (such as those in healthcare).

\printbibliography
\newpage
\onecolumn
\appendix
\section{Proofs.}

\subsection{Proof of proposition \ref{prop:prop0}}

We can provide a relatively straightforward proof using the direct definition of the CDF's $F_{ij}$ and $G_{ij}$:

\begin{align*}
    \L(y,1) = \mathbb{P}(y_{ij} = y, d_{ij}=1)&=\mathbb{P}\left(t_{ij}\in[y,y+\psi) , \psi.\lfloor \frac{t_{ij}}{\psi}\rfloor<\psi.\lfloor \frac{\varsigma_{ij}}{\psi}\rfloor\right) \\
    &=\P\left(t_{ij}\in[y,y+\psi), \varsigma_{ij}\geq \psi\cdot\lfloor \frac{t_{ij}}{\psi}\rfloor + \psi \right)\\
    &= \int_{y}^{y+\psi}\left(1-G_{ij}\left(\psi.\lfloor\frac{t}{\psi}\rfloor+\psi\right)\right)dF_{ij}(t)\\ \size{.6}{\text{since } \psi.\lfloor\frac{t}{\psi}\rfloor+\psi = y+\psi\text{ for any }t\in[y,y+\psi)} \quad\quad&=\left(1-G_{ij}(y+\psi)\right)\int_{y}^{y+\psi}dF_{ij}(t) \\
    &=\left(1-G_{ij}(y+\psi)\right).\left(F_{ij}(y+\psi)-F_{ij}(y)\right)
\end{align*}
\begin{align*}
    \L(y,0) = \mathbb{P}(y_{ij} = y, d_{ij}=0)&=\mathbb{P}\left(\varsigma_{ij}\in[y,y+\psi) , \psi.\lfloor \frac{t_{ij}}{\psi}\rfloor\geq\psi.\lfloor \frac{\varsigma_{ij}}{\psi}\rfloor\right) \\
    &=\P\left(\varsigma_{ij}\in[y,y+\psi), t_{ij}\geq \psi.\lfloor \frac{\varsigma_{ij}}{\psi}\rfloor\right)\\
    &= \int_{y}^{y+\psi}\left(1-F_{ij}\left(\psi.\lfloor\frac{\varsigma}{\psi}\rfloor \right)\right)dG_{ij}(\varsigma)\\ 
    \size{.6}{\text{since } \psi.\lfloor\frac{\varsigma}{\psi}\rfloor = y\text{ for any }\varsigma\in[y,y+\psi)} \quad\quad &=\left(1-F(y)\right)\int_{y}^{y+\psi}dG(\varsigma) \\
    &=\left(1-F(y)\right).\left(G(y+\psi)-G(y)\right)
\end{align*}

\subsection{Proof of Propositions \ref{prop:prop1}--\ref{prop:prop3}}

    The proof is trivial for \ref{prop:prop1} since setting $d_{ij}=0$ shrinks down the exact distribution of $\left(\nu_i\big|y_{ij},d_{ij}=0\right)$ in \ref{eq:4} to $\Gamma\size{.8}{\left(\alpha,\xi_{ij}\right)}$.
    
    For \ref{prop:prop2}, when $\epsilon_{ij}\to0$, we rewrite \ref{eq:4} in terms of $\epsilon_{ij}$:
    $$\Gamma\left(\nu_i\big|y_{ij},d_{ij}=0\right)=\frac{\nu_i^{\alpha-1}\cdot\exp\left(-\nu_i\cdot\left(\frac{\xi_{ij}+\xi_{ij}'}{2}\right)\right)}{\Gamma(\alpha)\cdot\left(\frac{\xi_{ij}+\xi_{ij}'}{2}\right)^{\!\!-\alpha}}\cdot \mathbf H(\epsilon_{ij},\nu_i)$$

    where:
    $$\mathbf H(\epsilon_{ij},\nu_i):=\frac{\exp\left(\nu_i\cdot\left(\frac{\xi_{ij}\cdot\epsilon_{ij}}{2}\right)\right)-\exp\left(-\nu_i\cdot\left(\frac{\xi_{ij}\cdot\epsilon_{ij}}{2}\right)\right)}{\left(\frac{2}{2+\epsilon_{ij}}\right)^{-\alpha}-\left(\frac{2+2\epsilon_{ij}}{2+\epsilon_{ij}}\right)^{-\alpha}}$$

    Expanding $\mathbf H(\epsilon_{ij},\nu_i)$ around $\epsilon_{ij}\to 0$ gives us:
    $$\mathbf H(\epsilon_{ij},\nu_i)=\frac{\nu_i\cdot\xi_{ij}}{\alpha}\left(1+\frac{\epsilon_{ij}}{2}\right)+\mathcal{O}(\epsilon^2)=\frac{\nu_i}{\alpha}\left(\frac{\xi_{ij}+\xi_{ij}'}{2}\right)+\mathcal{O}(\epsilon^2)$$ 

    Thus:
    $$\Gamma\left(\nu_i\big|y_{ij},d_{ij}\right)=\frac{\nu_i^{\alpha}\cdot\exp\left(-\nu_i\cdot\left(\frac{\xi_{ij}+\xi_{ij}'}{2}\right)\right)}{\Gamma(\alpha+1)\cdot\left(\frac{\xi_{ij}+\xi_{ij}'}{2}\right)^{\!\!-(\alpha+1)}}+\mathcal{O}(\epsilon^2)$$

    which converges to $\Gamma\size{.8}{(\alpha\!+\!1,\size{.8}{\frac{\xi_{ij}+\xi_{ij}'}{2}})}$ as $\epsilon_{ij}\to0$.

    Similarly, when $\epsilon_{ij}\to1$, equation \ref{eq:4} can be rewritten as:
    $$\Gamma\left(\nu_i\big|y_{ij},d_{ij}=0\right)=\frac{\nu_i^{\alpha-1}\cdot\exp\left(-\nu_i\cdot\xi_{ij}\right)}{\Gamma(\alpha)\cdot\xi_{ij}^{-\alpha}}\cdot\mathbf K(\epsilon_{ij},\nu_i)$$

    Where:
    $$\mathbf K(\epsilon_{ij},\nu_i):=\frac{1-\exp\left(-\nu_i\cdot\xi_{ij}\left(\frac{\epsilon_{ij}}{1-\epsilon_{ij}}\right)\right)}{1-\left(\frac{1}{1-\epsilon_{ij}}\right)^{-\alpha}}$$

    And the expansion of $\mathbf K(\epsilon_{ij},\nu_i)$ around $\epsilon_{ij}\to 1$ gives us:
    $$\mathbf K(\epsilon_{ij},\nu_i)=1+\mathcal{O}((1-\epsilon_{ij})^\alpha)$$

    Thus $\Gamma\left(\nu_i\big|y_{ij},d_{ij}\right)$ coverges to $\Gamma\size{.8}{\left(\alpha,\xi_{ij}\right)}$ as $\epsilon_{ij}\to1$.

    Finally, for \ref{prop:prop3}, we seek a gamma distribution that matches the posterior heterogeneity distribution in the first two moments:
    \begin{align*}
    \mu_1&=\mathbb{E}\left(\nu_i\big|y_{ij},d_{ij}\right)=\alpha\cdot\frac{\xi_{ij}^{-(\alpha+1)}-{\xi_{ij}'}^{-(\alpha+1)}}{\xi_{ij}^{-\alpha}-{\xi_{ij}'}^{-\alpha}} \\ \mu_2&=\mathbb{E}\left(\nu_i\big|y_{ij},d_{ij}\right)=\left(\alpha^2+\alpha\right)\cdot\frac{\xi_{ij}^{-(\alpha+2)}-{\xi_{ij}'}^{-(\alpha+2)}}{\xi_{ij}^{-(\alpha+1)}-{\xi_{ij}'}^{-(\alpha+1)}}
    \end{align*}

    We solve for the parameters of the approximate distribution $\Gamma\size{.8}{(\tilde\alpha, \tilde\kappa)}$:
    $$\tilde\alpha = \frac{\mu_1^2}{\mu_2-\mu_1^2}\ \ \ \ ;\ \ \tilde\kappa = \frac{\mu_1}{\mu_2-\mu_1^2}$$

    The above is a system of equations which yields a unique solution when $\epsilon_{ij}\in(0,1)$ equal to the expressions given in Proposition \ref{prop:prop3}. Moreover, it can be shown that as $\epsilon_{ij}$ approaches its extreme values, the solution of the system converges to the gamma distribution specified in Proposition \ref{prop:prop2}.

\subsection{Proof of Theorem \ref{theo:0}}

We need to show that $y^* = {o}_{\mathcal{P}}(n\cdot\psi)$. First, assume that only one spell is observed from each individual. Since $y^*=\max\limits_{\size{.6}{i,j}}y_{ij}$, we can invoke the Order statistics theorem, which guarantees that for all $\epsilon>0$:
\begin{align*}
    \mathbb{P}(y^*\leq n\cdot\psi\cdot\epsilon)=& \prod_{i=1}^n\left(1-\mathbb{P}( y_{i1}> n\cdot\psi\cdot\epsilon)\right)\\ \size{.6}{\text{by Chebyshev's inequality}}\quad\quad\quad\quad
     \geq&\left(1-\frac{\mathbb{E}(y_{i1}^2)}{n^2\cdot\psi^2\cdot\epsilon^2}\right)^n\\ \size{.6}{\text{since } \E(y_{ij}^2)<\infty\text{ and }\psi^{-2} = o(n)}\quad\quad
     \xrightarrow[]{n\to\infty}& \lim_{n\to\infty} \left(1-\frac{\E(y_{i1}^2)}{n\cdot \psi^2\cdot \epsilon}\right)=1
\end{align*}

Next, for any finite number of spells, $J$, we use the law of total expectation to conclude:
\begin{align*}
    \mathbb{P}(y^*\leq n\cdot\psi\cdot\epsilon)=& \prod_{i=1}^n\left(1-\mathbb{P}( \{ y_{ij}\}_{j=1}^J> n\cdot\psi\cdot\epsilon)\right)\\ \size{.6}{\text{since spells are independent conditional on } \nu_i}\quad\quad
    =& \prod_{i=1}^n \size{.8}{\bigg(}1-\E\size{.7}{\bigg(}\prod_{j=1}^J \P\big(y_{ij}>n\cdot\psi\cdot\epsilon\big|\nu_i\big)\size{.7}{\bigg)}\size{.8}{\bigg)}\\ \size{.6}{\text{by Chebyshev's inequality}}\quad\quad
    \geq& \left(1-\frac{\E\left(\E\left(y_{i1}^2\big|\nu_i\right)^J\right)}{(n^{2}\cdot\psi^2\cdot\epsilon^2)^J}\right)^n\\
    \xrightarrow[]{n\to\infty}& \lim_{n\to\infty} \left(1-\frac{\E\left(\E\left(y_{i1}^2\big|\nu_i\right)^J\right)}{n^{2J-1}\cdot \psi^{2J}\cdot \epsilon^{2J}}\right)\\
    \size{.6}{\text{by Jensen's inequality}}\quad\quad&\geq \lim_{n\to\infty} \left(1-\frac{\E\left(y_{i1}^{2J}\right)}{n^{2J-1}\cdot \psi^{2J}\cdot \epsilon^{2J}}\right)\\
    \size{.6}{\text{since } \E(y_{ij}^{2J})<\infty\text{ and }\psi^{-2} = o(n)}\quad\quad&=1
\end{align*}
Lastly, we have that: $$\psi^{-2} = o(n) \xrightarrow[]{\size{.6}{J\!>\!1}\ \to\ \size{.6}{2J\!-\!1\!>\!J}}\psi^{-2J} = o(n^J) = o(n^{2J-1})$$

Combining the two results leads to the conclusion that:

\begin{align*}
    \lim_{n\to\infty} \left(1-\frac{\E\left(\E\left(y_{i1}^2\big|\nu_i\right)^J\right)}{n^{2J-1}\cdot \psi^{2J}\cdot \epsilon^{2J}}\right) = 1 \\
\end{align*}

Which proves the theorem in the general case.

\subsection{Proof of Theorem \ref{theo:2}}

Denote by $\widetilde{\mathcal S}\coloneq \left\{k\cdot\psi\big|k\in\mathbb N_0\right\}\times\{0,1\}\times \mathcal S(\mathbf x)$ the space of possible values for the triplet $[y_{ij}, d_{ij},\mathbf x_{ij}]$, and by $\Theta = \big\{\Phi(.),\{\delta_1,\dots,\delta_r\},\alpha,\kappa\big\}$ the set of model parameters. Define a function $f_{\Theta}:\bigcup\limits_{\size{.55}{J\in\mathbb N}}\widetilde{\mathcal S}^{J}\to\mathbb R$ that maps each dataset of the the form $\left[\mathbf y_{i},\mathbf d_{i},\mathbf X_{i}\right]=\left\{[y_{i1}, d_{i1},\mathbf x_{i1}],\dots,[y_{iJ}, d_{iJ},\mathbf x_{iJ}]\right\}$, to its corresponding panel likelihood $\boldscr L_i\big(\mathbf y_i, \mathbf d_i\big| \Theta\big)$ as defined in equation \ref{eq:2.5}. Let $\widetilde{\mathcal S}_\Theta\coloneq \left\{\size{.8}{\bigg(}\left[\mathbf y,\mathbf d,\mathbf X\right], f_\Theta(\mathbf y,\mathbf d,\mathbf X)\size{.8}{\bigg)}\bigg|\left[\mathbf y,\mathbf d,\mathbf X\right]\in\bigcup\limits_{\size{.6}{J\in\mathbb N}}\widetilde{\mathcal S}^J\right\}$ denote the graph of the function $f_\Theta$. To prove unique identifiability, we need to show that the mapping $T:\Theta\mapsto \widetilde{\mathcal S}_\Theta$ between the space of all possible parameters and the space of all possible graphs is one-to-one. Note, that $T$ maps each parameter to an entire space of the pairs of ${\big(}\left[\mathbf y,\mathbf d,\mathbf X\right], f_\Theta(\mathbf y,\mathbf d,\mathbf X){\big)}$. Hence, to show that the mapping is one-to-one, it suffices to show one element of $\widetilde {\mathcal S}_{\Theta}$ that is not in $\widetilde {\mathcal S}_{\Theta'}$ for any $\Theta\neq \Theta'$.

Accordingly, let $J=2$ and consider the two datasets $[\mathbf y_1, \mathbf d_1, \mathbf X_1]=\left\{[y_{11},d_{11},\mathbf x_{11}],[y_{12},d_{12},\mathbf x_{11}] \right\}$ and $[\mathbf y_2, \mathbf d_2, \mathbf X_2]=\left\{[y_{21},d_{21},\mathbf x_{21}],[y_{22},d_{22},\mathbf x_{22}] \right\}$, where $d_{11}\!=\!d_{12}\!=\!d_{21}\!=\!d_{22}=1$, $y_{11}\!=\!y_{12}\!=\!y_{21}\!=\!y_{22}=0$. Moreover, Leveraging assumption \ref{as:5}, let $\Phi(x_{11}) = \Phi(x_{12})=\Phi(x_{21}) = 1$ while $\Phi(x_{22})\neq 1$. Accordingly, we would have the following likelihood ratio between the two datasets:
\begin{align*}
    \frac{\boldscr L_1(\mathbf y_1, \mathbf d_1)}{\boldscr L_2(\mathbf y_2, \mathbf d_2)} &= \frac{\size{.85}{\left(1-\bigg(1+\frac{\delta_1}{\kappa}\bigg)^{\!\!-\alpha}\right)\cdot \left(1-\bigg(1+\frac{\delta_1}{\tilde\kappa}\bigg)^{\!\!-\tilde\alpha}\right)}}{\size{.85}{\left(1-\bigg(1+\frac{\delta_1}{\kappa}\bigg)^{\!\!-\alpha}\right)\cdot \left(1-\bigg(1+\frac{ \Phi(\mathbf{x}_{ij})\cdot \delta_1}{\tilde\kappa}\bigg)^{\!\!-\tilde\alpha}\right)}}\\
    &= \frac{\size{.85}{1-\bigg(1+\frac{\delta_1}{\tilde\kappa}\bigg)^{\!\!-\tilde\alpha}}}{\size{.85}{1-\bigg(1+\frac{ \Phi(\mathbf{x}_{ij})\cdot \delta_1}{\tilde\kappa}\bigg)^{\!\!-\tilde\alpha}}}
\end{align*}
where
\begin{align*}
    &\tilde\alpha=\alpha\ \cdot\!\size{.8}{\left({\frac{(1-(\frac{\kappa}{\kappa + \delta_1})^{\alpha+1})^2}{(1-(\frac{\kappa}{\kappa + \delta_1})^{\alpha})(1-(\frac{\kappa}{\kappa + \delta_1})^{\alpha+2})-\alpha\cdot(\frac{\kappa}{\kappa + \delta_1})^\alpha(\frac{\delta_1}{\kappa + \delta_1})^2}}\right)}\quad ; \quad
    \tilde\kappa={\xi_{ij}\!\cdot\!\size{.8}{\left(\frac{(1-(\frac{\kappa}{\kappa + \delta_1})^{\alpha})(1-(\frac{\kappa}{\kappa + \delta_1})^{\alpha+1})}{(1-(\frac{\kappa}{\kappa + \delta_1})^{\alpha})(1-(\frac{\kappa}{\kappa + \delta_1})^{\alpha+2})-\alpha\cdot(\frac{\kappa}{\kappa + \delta_1})^\alpha(\frac{\delta_1}{\kappa + \delta_1})^2}\right)}}
\end{align*}
are derived according to proposition \ref{prop:prop3}. It is readily observable that this equation provides a unique expression for $\Phi(\mathbf x_{22})$. Moreover, the left-hand-side of the equation is strictly monotone in terms $\{\delta, \alpha,\kappa\}$; therefore, the identifiability of the parameters is concluded.


\end{document}